\documentclass[twocolumn,groupedaddress,showpacs,floatfix,notitlepage,color]{revtex4-1}
\usepackage{graphicx}
\usepackage{epsfig}
\usepackage{amsfonts}
\usepackage{color}
\usepackage{ulem}
\usepackage{amsmath}
\usepackage{mathrsfs}
\usepackage{subfigure}
\usepackage[autostyle]{csquotes}
\usepackage[percent]{overpic}
\usepackage{pict2e}
\usepackage[retainorgcmds]{IEEEtrantools} 

\begin{document}

\title{Temporal dynamics in immunological synapse: Role of thermal fluctuations in signaling}

\author{Daniel R Bush}
\affiliation{                    
Non-linearity and Complexity Research Group - Aston University, Aston Triangle, Birmingham, B4 7ET, UK}
\author{Amit K Chattopadhyay}
\affiliation{                    
Non-linearity and Complexity Research Group - Aston University, Aston Triangle, Birmingham, B4 7ET, UK}
\email{a.k.chattopadhyay@aston.ac.uk}
\begin{abstract}
The article analyzes the contribution of {\it stochastic thermal fluctuations} in the attachment times of the immature T-cell receptor TCR: peptide-major-histocompatibility-complex pMHC immunological synapse bond. The key question addressed here is the following: how does a synapse bond remain stabilized in the presence of high frequency thermal noise that potentially equates to a strong detaching force? Focusing on the average time persistence of an immature synapse, we show that the high frequency nodes accompanying large fluctuations are counterbalanced by low frequency nodes that evolve over longer time periods, eventually leading to signaling of the immunological synapse bond primarily decided by nodes of the latter type. Our analysis shows that such a counterintuitive behavior could be easily explained from the fact that the survival probability distribution is governed by two distinct phases, corresponding to two separate time exponents, for the two different time regimes. The relatively shorter time scales correspond to the cohesion:adhesion induced immature bond formation whereas the larger time reciprocates the association:dissociation regime leading to TCR:pMHC signaling. From an estimation of the bond survival probability, we show that at shorter time scales, this probability $P_{\Delta}(\tau)$ scales with time $\tau$ as an universal function of a rescaled noise amplitude $\frac{D}{\Delta^2}$, such that $P_{\Delta}(\tau)\sim \tau^{-(\frac{\Delta}{\sqrt{D}}+\frac{1}{2})}$, $\Delta$ being the distance from the mean inter-membrane (T cell:Antigen Presenting Cell) separation distance. The crossover from this shorter to a longer time regime leads to an universality in the dynamics, at which point the survival probability shows a different power-law scaling compared to the one at shorter time scales. In biological terms, such a crossover indicates that the TCR:pMHC bond has a survival probability with a slower decay rate than the longer LFA-1:ICAM-1 bond justifying its stability. 
\end{abstract}
\date{\today}


\maketitle

\section{Introduction}
Interactions between immune cells (T cells) and antigen presenting cells (APCs) are fundamental to the activation of an adaptive immune response.
Cell to cell contact enables protein complexes on the opposing membranes to come in \enquote{close contact} with each other, facilitating bonding between them.
Integrin-ligand pairs form bonds producing conformational changes on the intracellular portion of the membrane bound proteins, whereupon signals are carried through intracellular signaling pathways.

A necessary bond for T cell activation is the one formed between the T cell receptor (TCR) and the major-histocompatibility complex molecule with bound antigenic peptide (pMHC).
During the early stages of the T-cell lifecycle, the TCR is primed to recognise particular peptides of previously encountered antigenic material.
MHC molecules on the surface of APCs contain bound antigenic peptides and when the affinity between the TCR and pMHC are favourable, a bond is formed with an approximate length of 15 nm.
The intracellular conformational changes in the bound TCR facilitate Src kinase signal transduction that ultimately lead to cell proliferation and activation of the immune function~\cite{bib:springer_2012,bib:bunnell_2002,bib:janeways_2008}.

Another bond formed during the initial cell to cell contact is that between the intercellular adhesion molecule-1 (ICAM-1) and the leukocyte function associated-1 adhesion molecule (LFA-1).
The ICAM-1:LFA-1 bond length is approximately $45$ nm, significantly larger than the TCR:pMHC bond.
During the initial stages of cell contact the larger bonds (ICAM-1:LFA-1) localise at the centre of the contact zone, with small patches of TCR:pMHC bonds forming at the edge of the central zone.
Fluorescent tagging shows heterogenous segregation and aggregation of the molecules in the contact interface, attributed to the different bond length scales~\cite{bib:bunnell_2010,bib:davis_2006}, that leads to the formation of an immunological synapse (IS).

During a TCR's engagement with a pMHC molecule, the Src-family protein tyrosine kinases Lck and Fyn phosphorylate and activate a number of complexes  (ZAP-70, SLP-76, LAT) that are recruited to the immunoreceptor tyrosine-based activation motifs (ITAM) on the TCR$\zeta$ chain.
The recruited complexes assemble to transmit the signal through phosphorylation and activation of downstream signaling complexes.
This leads to the activation of transcription factors in the nucleus, initiating gene transcription.
A precursor to gene transcription is the elevation of cytoplasmic Ca$^{2+}$ concentrations measured to peak around 12 seconds~\cite{bib:bunnell_2002,bib:rotnes_1994}, minutes before the mature IS forms.

Previous works \cite{bib:chattopadhyay_2007,bib:bush_2014} focused on the average time persistence of the bond duration, studying a range of bond lengths consistent with the TCR:pMHC and ICAM-1:LFA-1 bonds~\cite{bib:bush_2014}. This current work builds on the same membrane interaction model to investigate the role of extremal value statistics in the temporal dynamics of the immunological synapse process. More specifically, we want to analyze the contribution of the extremal time dynamical nodes in arriving at the 2-4.5 seconds' premature synapse bonding time that was previously estimated \cite{bib:bush_2014}. In a remarkable departure from uneducated expectation, we show that although the larger nodes dissipate more membrane energy through faster hydrodynamic dissipation, the overall statistics is only sparingly affected by these contributions. Rather the persistence profile is determined by perturbations with smaller amplitudes. In order to explain this finding, we have calculated the decay rate of the bond survival probability under a range of thermal noise strengths to understand the importance of energy dissipation and the corresponding rate of dissipation, thereby to attain a limiting parametric description of our stochastically forced linearly stable model.

The article is organized as follows. 
In section~\ref{sec:model} we present a description of the stochastically forced membrane model.  
Section~\ref{sec:sim_method} then details the numerical analysis, including the close contact survival statistics algorithm and its implementation. 
This is followed by an analysis of the numerical results in section~\ref{sec:results} that is subdivided into three parts, the first of which analyzes the small time phase, subsection~\ref{sec:results:large_t_phase} analyzes large time phase statistics and section~\ref{sec:results:extremal_value_statistics} focuses on extremal value statistics, including comparisons with subsections~\ref{sec:results:small_t_phase} and~\ref{sec:results:large_t_phase}. This is followed by a conclusion and future projections.

\section{The Model}
\label{sec:model}
\subsection{The TCR:APC Membrane Fluctuation Model}
We analyze the qualitative dynamics of the cell-cell separation distance using a linear model, derived from a linear stability analysis of the nonlinear SA model~\cite{bib:bush_2014,bib:qi_2001}.
The equation of motion for the separation distance, $\phi(\mathbf{x},t)$, at a given point $\mathbf{x}$ on the membrane surface is given by
\noindent
\begin{equation}
M \frac{\partial \phi}{\partial t} = -B \nabla^4 \phi + \gamma \nabla^2 \phi - \lambda \phi + \eta,
\label{eqn:linear_model}
\end{equation}
\noindent
where $B$ is the coefficient of the membrane rigidity, $\gamma$ is the surface tension, $\lambda$ quantifies the linearized relaxation kinetics close to equilibrium and $M$ is the membrane damping (phenomenological) constant.
As in standard membrane dynamics, the membrane rigidity term and the surface relaxation terms create a force balance by working against each other while the contribution from the surrounding coreceptor molecules is encapsulated in the linear $-\lambda \phi$ term.
The thermal noise $\eta(\mathbf{x},t)$ is assumed to be Gaussian white noise defined through fluctuation-dissipation kinetics~\cite{bib:chattopadhyay_2007}
\begin{IEEEeqnarray}{rCl}
\left< \eta \left( \mathbf{x},t \right) \right>  & = & 0
 \IEEEyessubnumber \label{noise_1} \\
\left< \eta \left( \mathbf{x},t \right) ~ \eta \left( \mathbf{x}',t' \right) \right>  & = & 2 D~{\small k}_B{\footnotesize T} \delta \left( \mathbf{x} - \mathbf{x}' \right) \delta \left( t-t' \right)
 \IEEEyessubnumber \label{noise_2} \qquad
\end{IEEEeqnarray}
The range of validity of this model is limited to the start of the immunological synapse patterning and does not describe the dynamics that lead to the self-organization of the mature synapse \cite{bib:bush_2014}.

\subsection{The Single Threshold Model}
\begin{figure}[htbp!]
	\centering
	\subfigure[]{
		\includegraphics[width=0.48\textwidth]{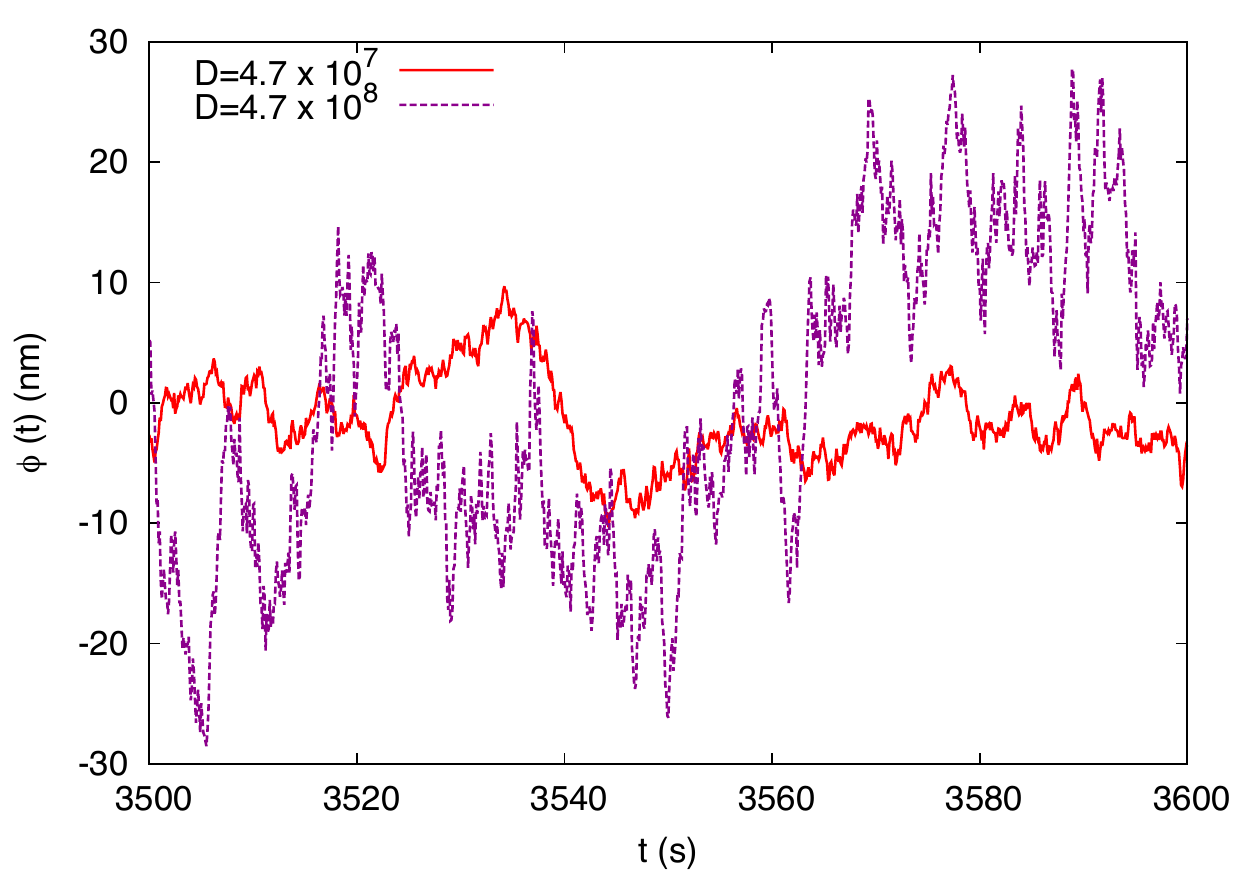}
		\label{fig:phi_solution}		
	} 
	\\
	\subfigure[]{
		\includegraphics[width=0.45\textwidth]{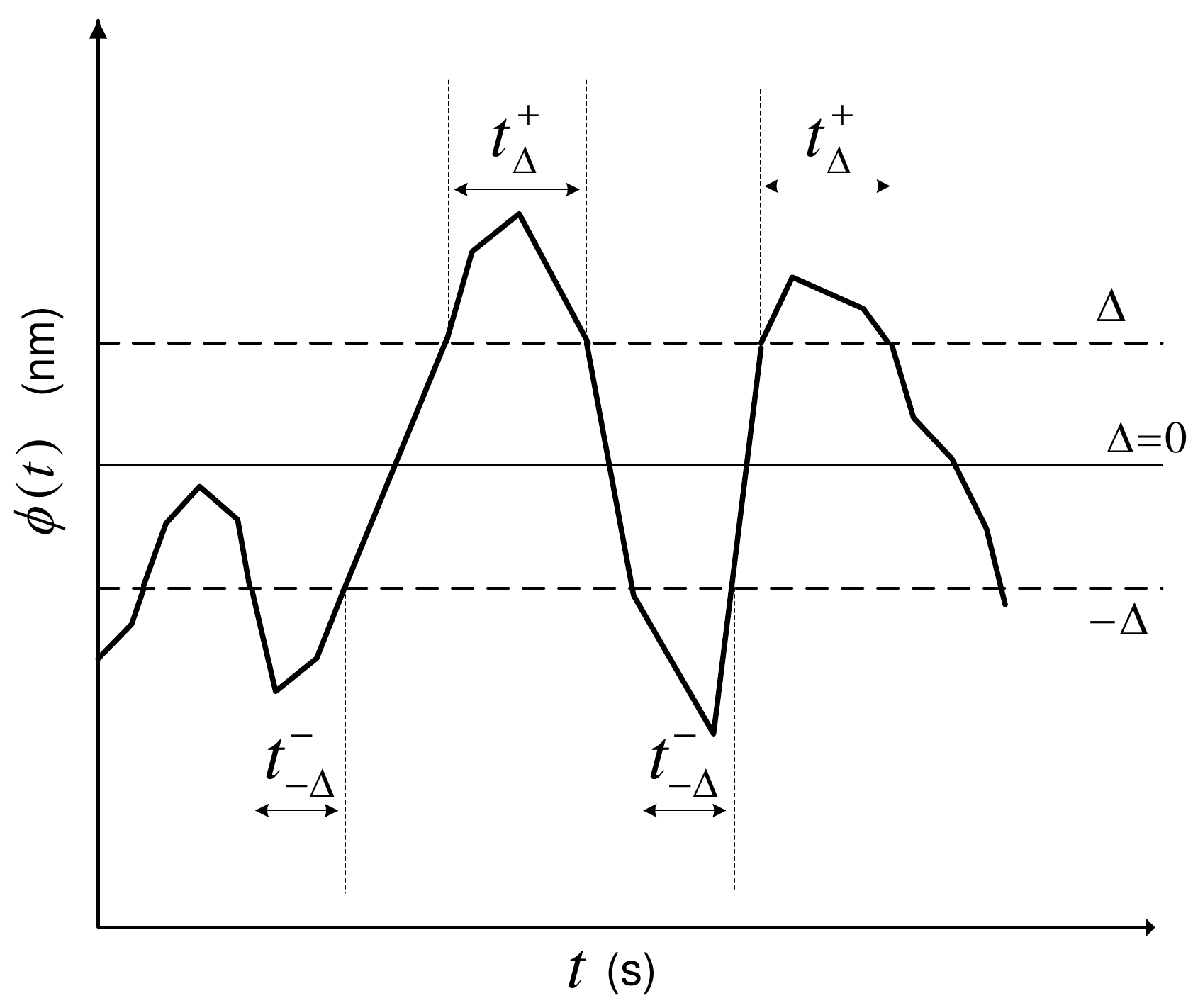}
		\label{fig:graph_method_single_threshold}
	}
	\caption[(Color online) The simulated solution for the separation distance and time persistence definition]{(Color online) \textbf{Examples of simulated solutions for the separation distance and time persistence definition.} Figure~\ref{fig:phi_solution} shows the simulated time dependent solution for $\phi$ using two different delta correlated noise strengths and~\ref{fig:graph_method_single_threshold} shows a schematic illustration of the $t^-_{-\Delta}$ and $t^+_{\Delta}$ regions that represent persistent bonds with a bond length $\Delta$.}
\end{figure}

There are two distinct length scales of separation between participating membranes in the IS problem that range from 15 nm (TCR:pMHC) to 45 nm (integrin:ligand)~\cite{bib:burroughs_2002}.
Equation~\eqref{eqn:linear_model} describes the local fluctuations about a mean separation distance between the membranes.
We introduce a single threshold value $\Delta$, that defines a distance from the mean separation distance that may be used to analyze the dynamics away from the mean separation distance.
Two opposing membranes are said to be within a close contact distance if the separation distance is less than $-\Delta$nm~\cite{bib:chattopadhyay_2007}, such that if the mean separation distance were 25nm, then $\Delta=10$nm would describe a close contact definition of 15nm.
The time persistence of a bond is given by the length of time the separation distance remains below this threshold value.
The solution to equation~\eqref{eqn:linear_model} is the stochastic variable $\phi(\mathbf{x},t)$ that is a {\it Gaussian Stationary Process} (GSP) fluctuating about $\phi=0$~\cite{bib:bush_2014}.
Figure~\ref{fig:phi_solution} shows sample simulation results for two different values of the noise strength, $D$. The results indicate that an increase in noise strength relates to a GSP with a larger amplitude.

A schematic illustration of the method used to determine the bond time persistence instances is shown in figure~\ref{fig:graph_method_single_threshold}.
The $t^-_{-\Delta}$ regions correspond to the length of time a bond (with length $\Delta$ from the mean separation distance) persists and are calculated as the time $\phi$ spends below the $-\Delta$ threshold.
$\phi$ is statistically symmetrical about $\Delta=0$, but for $\Delta \neq 0$ the symmetry in the $\phi$ direction is broken, whereupon the statistics below $-\Delta$ are different from the statistics above $-\Delta$.
In this case, the persistence above $-\Delta$ would indicate the time duration where no bonding is taking place.
However, in accordance with the statistical symmetry of the GSP about $\Delta=0$, the statistics for $\phi < -\Delta$ are equivalent to the statistics for $\phi > \Delta$.
A $t^+_{\Delta}$ region is defined as any time interval between two successive crossings of the $\phi=\Delta$ line from below, to the time taken to return below the $\Delta$ threshold.
In this case each instance of time persistence is given by $t^+_{\Delta} = t_2 - t_1$, where
\begin{IEEEeqnarray}{rCl}
& & \phi(t_1) = \phi(t_2) = \Delta \nonumber \\
& & \phi(t) \geq \Delta, \qquad t_1 < t < t_2
\label{eqn:time_persistence}
\end{IEEEeqnarray}
The different instances of time persistence are assumed to be statistically independent ({\it independent interval approximation}~\cite{bib:bray_2013,bib:majumdar_1996b}), then the average time persistence is given by
\begin{equation}
<t^+_\Delta> = \frac{1}{T} \sum_{i=1}^T t^{+ (i)}_{\Delta}
\label{eqn:t_average}
\end{equation}
where $t^{+ (i)}_{\Delta}$ is the $i$th instance of time persistence above $\Delta$ and $T$ is the total number of instances.
In the following analysis we drop the `+' for ease of notation, where it is assumed $t_\Delta$ is the time persistence above $\Delta$.

A major emphasis of this study is to analyze the effect of extremal values on the time persistence calculated using this statistical average technique as has often been shown to be of great importance in understanding the expected variation between the theoretical analysis presented and analogous biological experiments \cite{bib:chattopadhyay_2007,bib:bush_2014}. As an example of the latter type, quite often in the nanospectroscopy of flagellar (e.g. sperm \cite{bib:hilfinger_2009,bib:bayly_2014}) dynamics, ensemble averaging is a serious issue due to the perceived lack of ergodicity in such dynamics.  Technically, what this will imply is an understanding of the role of the long tail in the $P_{\Delta}(\tau)$ probability distribution profile, as defined in equation (\ref{eqn:survival_probability}) later. We will see that high frequency nodes, the generator of extremal value statistics, surprisingly return negligible contribution thereby defining a \enquote{null hypothesis} of sorts. Many quantitative biological experiments and conclusions are based on singular or at best only a small number of observations. Such lack of statistical information implies that existing probabilistic theories, including previous immunological synapse based models, will be inadequate in dealing with such eventualities. A corollary of our present work is the development of a methodology to avoid having to explicitly deal with statistically large datasets, since at least for immature immunological synapse dynamics, our results clearly indicate that large amplitude fluctuations can be largely neglected in the statistical analysis, thereby limiting the available configuration space to a much smaller size than it would be otherwise. An even greater impact of this result will be evident in the future nonlinear modelling of the mature synapse model that has a much larger parametric space, accompanied with large amplitude fluctuations that we can neglect as a first approximation based on this present result.

\section{Simulation Method}
\label{sec:sim_method}
\subsection{Time Evolution of the Langevin Equation}
\label{sec:time_evolution_langevin}
We solved equation~\eqref{eqn:linear_model} in discretized time and space both for d=1+1 and d=2+1 dimensions.
We simulate for $\phi(x_i, t_n)$, where $x_i = i \Delta x$ and $t_n = n \Delta t$ with $n=0,1, \dots, N-1$ and $i=0,1,\dots ,L-1$.
Periodic boundary conditions are used with $N=10^5$ and $L=100$.
The following description covers the d=1+1 case, but can be easily extended to the d=2+1 case by using the appropriate spatial derivatives.

We compute the solution for $\phi_{i,n}=\phi(x_i,t_n)$ using a forward Eulerian difference scheme 
\begin{equation}
\phi_{i,n+1}=\phi_{i,n}+\Delta t \frac{\partial \phi_{i,n}}{\partial t}
\end{equation}
with the spatial derivatives in the Langevin equation given by
\begin{IEEEeqnarray}{rCl}
\nabla^2 \phi_{i,n} & = & \frac{\phi_{i-1,n} - 2 \phi_{i,n} + \phi_{i+1,n}}{(\Delta x)^2} \\
\nabla^4 \phi_{i,n} & = & \frac{\phi_{i-2,n} -4 \phi_{i-1,n} + 6 \phi_{i,n} -4 \phi_{i+1,n} + \phi_{i+2,n}}{(\Delta x)^4} \nonumber \\
& & 
\end{IEEEeqnarray}
We used $\Delta x = 1$ and values of $\Delta t = 0.1$ and $\Delta t = 0.01$ to ensure the iteration of the deterministic portion of the Langevin equation is stable.
The noise was scaled accordingly such that the equation iterated was
\begin{IEEEeqnarray}{rCl}
\phi_{i,n+1} & = & \phi_{i,n} - \Delta t \frac{B}{M} \nabla^4 \phi_{i,n} + \Delta t \frac{\gamma}{M} \nabla^2 \phi_{i,n} \nonumber \\
& & - \Delta t \frac{\lambda}{M} \phi_{i,n} + \frac{\sqrt{2 D~{\small k}_B T\: \Delta t}}{M} \zeta(x_i,t_n)
\label{eqn:finite_difference}
\end{IEEEeqnarray}
where $\zeta(x_i,t_n)$ is a Gaussian distributed random number with zero mean and unit variance. The core structure relies on a version of Stratonovich~\cite{bib:risken_1989} calculus in order to avoid explicit multiplicative noise in the basic model, in conformity with most biological models \cite{bib:qi_2001,bib:burroughs_2002} of this genre.

\subsection{Close Contact Survival Statistics}
\label{sec:close_contact_survival_statistics}
The $t_\Delta$ statistics were gathered using the time evolution of each point $x_i$.
If $\phi (x_i,t_n)$ crosses the threshold from below then  the point $t_n$ is stored and retained until the separation distance crosses back over the threshold at some time $t_{n+m}$, then the time persistence $t_\Delta = m \Delta t$ is added to the statistics.
The instances of $t_{\Delta}$ are stored and used to produce frequency distributions and the ensemble average as defined in equation~\eqref{eqn:t_average}.

For large enough statistics, the normalized frequency distribution is equivalent to the probability density for $t_{\Delta}$.
Then the probability that the time persistence is equal to or less than $\tau$ is given by $P(t_\Delta \leq \tau)$, with $P(t_\Delta \leq \infty)=1$.
For discrete $P$, as is our case, the probability density can be expressed as a sum of $\delta$ functions
\begin{equation}
W_\Delta (\tau) = \displaystyle \sum_{m=1}^{\infty} p_m^\Delta \delta(\tau - \tau_m)
\label{eqn:probability_density}
\end{equation}
where $\tau_m$ are multiples of $\Delta t$ and $p_m^\Delta$ is the first passage probability to find the discrete value $\tau_m$ for a given $\Delta$ value~\cite{bib:krug_1997a}. 
The persistence probability is then identified through the first passage statistics of the fluctuating interface
\begin{IEEEeqnarray}{rCl}
\label{eqn:transient_persistence_probability}
P_{\Delta}(\tau) & = & 1 - \int_0^\tau W_\Delta (\tau') \text{d}\tau'
\label{eqn:survival_probability}
\end{IEEEeqnarray}

\begin{figure}[htbp!]
	\centering
	\includegraphics[width=0.49\textwidth]{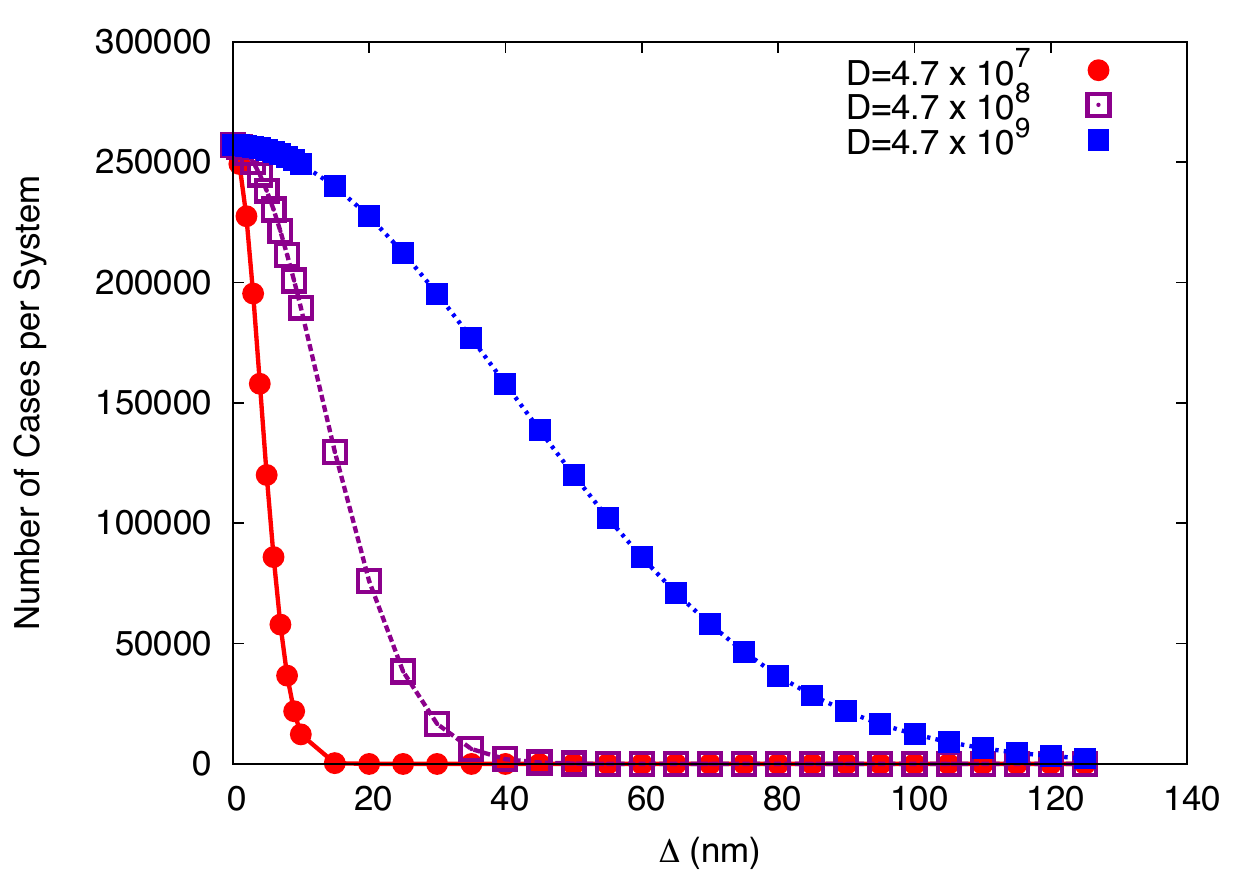}
	\caption[(Color online) Average number of cases per system run for $\text{d}=1+1$.]{(Color online) \textbf{Average number of cases per system run for $\text{d}=1+1$.}
	The number of statistics per simulated system is plotted against the $\Delta$ threshold value.
	Three thermal noise strengths are plotted to show the variation in the statistics as the noise strength changes.   }
	  \label{fig:average_cases_per_system}
\end{figure}

Using the system parameters described in section~\ref{sec:time_evolution_langevin}, a single d=1+1 system generates on average $2.5 \times 10^5$ statistics for $\Delta=0$.
We used an ensemble comprising $10^5$ systems, giving statistics in the order of $10^{10}$  for the ensemble calculations.
As the threshold increases the average number of statistics per system decreases as shown in figure~\ref{fig:average_cases_per_system}.
Increasing the thermal noise strength increases the range of $\Delta$ where statistics can be found, although the Gaussian profile is unchanged.
We also note the number of statistics is constant for $\Delta=0$, regardless of the noise strength.

Simulations for the d=2+1 case were performed on a lattice of  L=$50\times50$ and the number of statistics generated scales proportional to the number of spatial nodes.
The distribution of statistics is unaffected by the spatial dimension used (results tested on larger sized lattices too; conclusions remain the same), therefore we restrict our analysis to the d=1+1 case.

\section{Results}
\label{sec:results}

\begin{figure}[htbp!]
	\centering
	\includegraphics[width=0.48\textwidth]{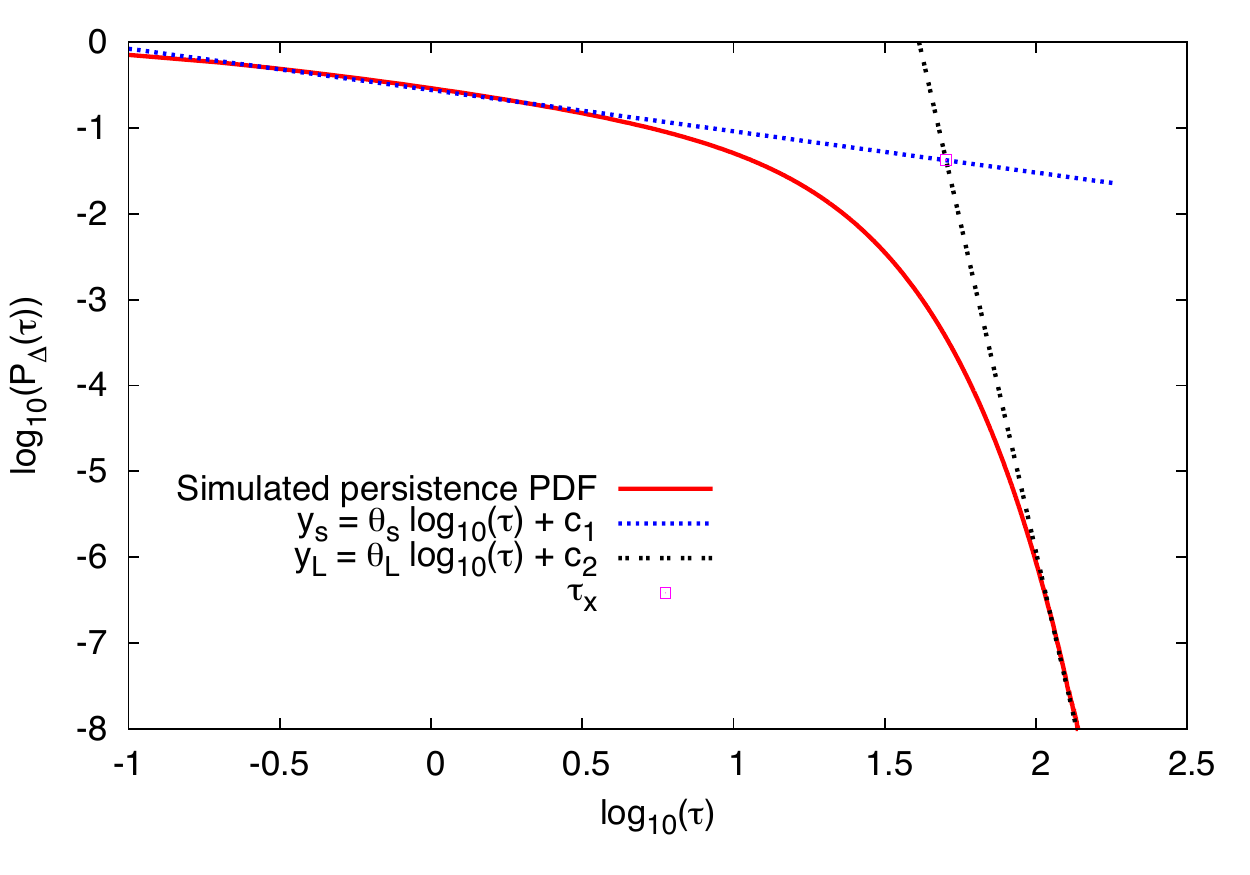}
	\caption[(Color online) Log:Log plot for the survival probability $P_{\Delta=0}(\tau)$]{(Color online) \textbf{Log:Log plot for the survival probability  $P_{\Delta=0}(\tau)$}.
	The gradients for the two fitted lines, $\theta_S$ and $\theta_L$, show the power law decay for the small $\tau$ and large $\tau$ time domains, respectively.
	The crossover point between these regimes is marked as the intersection between the fitted lines (square point)}
	\label{fig:log_pdf_function_method}
\end{figure}

For the IS problem we use coefficients $M=4.7\times 10^{6}~k_{B} T \mu \text{m}^{-4}$, $B=11.8~k_{B} T$, $\gamma=5650~k_{B} T \mu \text{m}^{-2}$ and $\lambda=6\times10^5~k_{B} T \mu \text{m}^{-4}$, with $D$ just large enough to stimulate the fluctuations, without dominating the dynamics~\cite{bib:chattopadhyay_2007,bib:bush_2014}.
Figure~\ref{fig:log_pdf_function_method} shows the log:log plot (solid line) for the persistence probability, $P_{\Delta}(\tau)$, from a numerical simulation for the parameter value $\Delta=0$ and the model parameter values mentioned above.
We observe two distinct linear regions corresponding to two different power law decay time exponents.
Using a least squares' linear fit in the log:log regime we find approximations for the time persistence exponents in each regime.
The transition region separating the two different scaling regimes has a length scale in the order of tens of seconds, with the crossover point $\tau_\times$ shown as the intersection between the least square fitted lines.
We define $\theta_S$ as the time persistence exponent for the small $\tau$ regime and $\theta_L$ as the exponent for the large $\tau$ regime.
\begin{IEEEeqnarray}{rCl}
P_{\Delta}(\tau) & \sim & \tau^{-\theta_S} \qquad \tau \ll \tau_\times \\
P_{\Delta}(\tau) & \sim & \tau^{-\theta_L} \qquad \tau \gg \tau_\times
\end{IEEEeqnarray}

\noindent
As the results clearly show, the system shows two different relaxation time scales, one dominated by diffusion and the other by the chemical force interacting with the stochastic forcing. 
In a way, this is complementary to the two time scale problem that was analyzed earlier \cite{bib:burroughs_2002}. 
In the following subsections we analyze the small and large $\tau$ regimes, with respect to the bond length and thermal noise strength.
And finally, we use the persistence probability density to understand the effect of high frequency fluctuations on the average statistics. 

\subsection{The Small $\tau$ Phase: The \enquote{Thermal Fluctuations} regime}
\label{sec:results:small_t_phase}
The small $\tau$ phase represents the regime where receptor:ligand complexes are associating and dissociating rapidly due to the impact of high frequency thermal fluctuations, within a relative small time space (measured by smaller number of time steps).
The bond duration in the small $\tau$ regime is not deemed sufficient to coincide with the elevation in intracellular Ca$^{2+}$ levels at 12 seconds~\cite{bib:bunnell_2002}.

However, this regime is statistically relevant due to the impact on the average time persistence.
The probability density associated with $\Delta=0$ inidicates approximately $30\%$ of statistical cases will persist for a single time step, whereupon they return below the threshold.
Using the notation from equation~\eqref{eqn:probability_density}, the first passage probability values are $p^0_1 = 0.29306$, $p^0_2 = 0.12993$ and $p^0_3 = 0.07751$, meaning $50\%$ of the $t_{\Delta=0}$ instances survive for three (or less) discrete time lengths.

\begin{figure}[htbp!]
	\centering
	\subfigure[]{
		\includegraphics[width=0.45\textwidth]{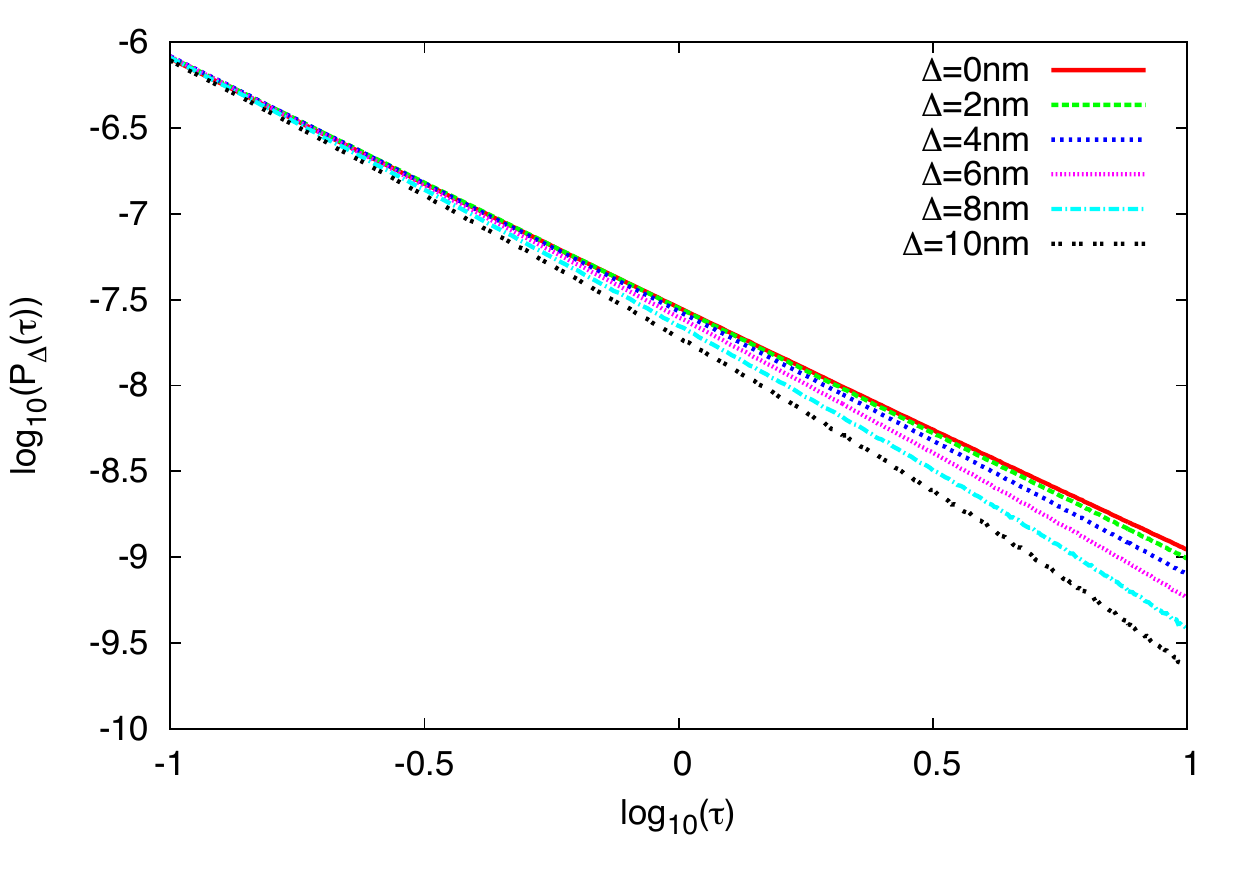}
		\label{fig:small_t_exponent_comparison}
	} 
	\\
	\subfigure[]{
		\includegraphics[width=0.45\textwidth]{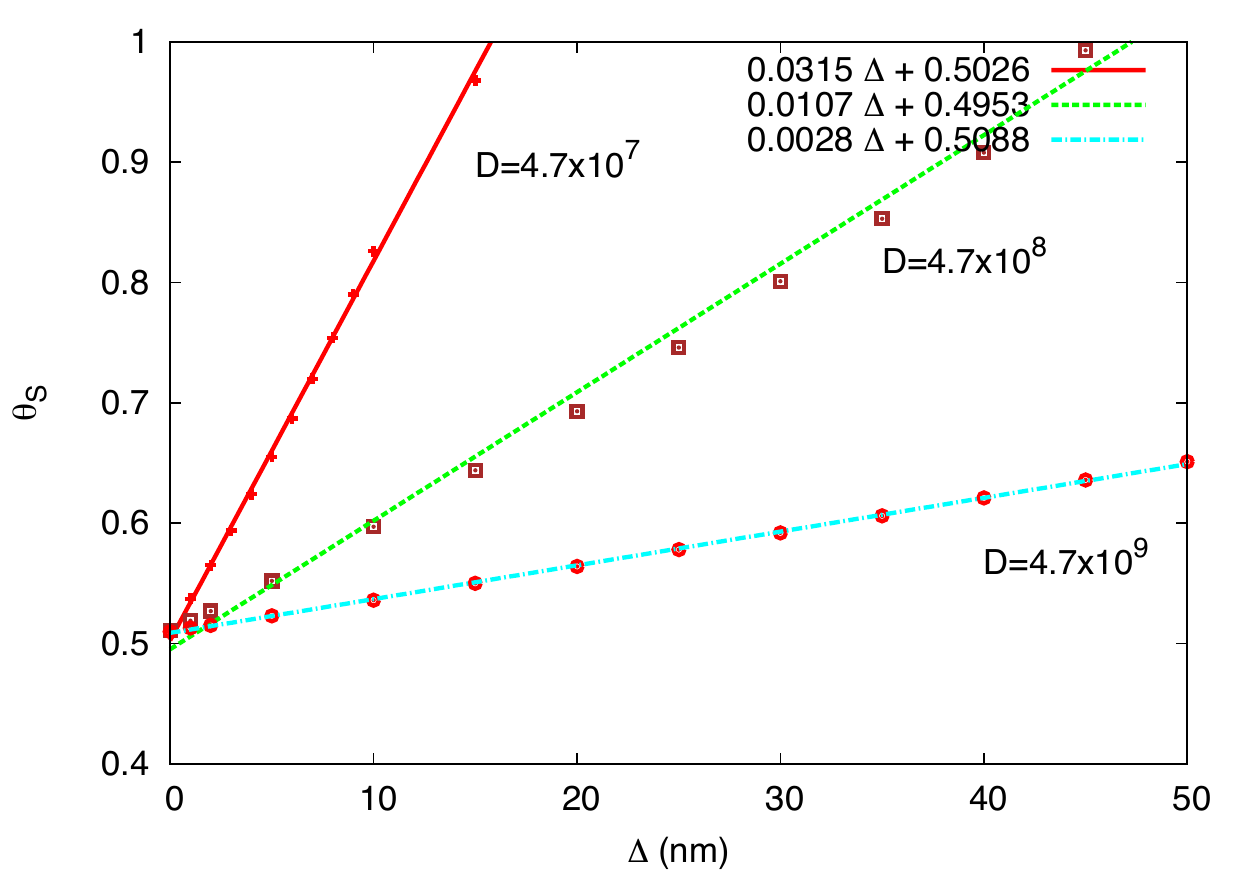}
		\label{fig:thetas_vs_delta}
	}
	\caption[(Color online) The small $\tau$ regime time exponent.]{(Color online) \textbf{The time exponent for the small $\tau$ regime.} 
	Figure~\ref{fig:small_t_exponent_comparison} shows the plot of the survival probability against time, in the log:log scale, for a range of $\Delta$ and 
	figure~\ref{fig:thetas_vs_delta} shows the time exponent $\theta_S$ plotted against $\Delta$ for three different thermal noise strengths.}
\end{figure}

Simulations using $\Delta t=0.01$ and $N=10^6$ were run and figure~\ref{fig:small_t_exponent_comparison} shows the log:log plot for the survival probability against time for different $\Delta$ in the small $\tau$ regime.
$\theta_S$ becomes steeper with increasing $\Delta$, indicating the rate of decay for the survival probability (and hence the survival probability) is dependent on $\Delta$.
The relationship between $\theta_S$ and $\Delta$ is shown in figure~\ref{fig:thetas_vs_delta} for a range of thermal fluctuation strengths.
In each case a near linear relationship exists between the time exponent and $\Delta$.
All three values of $D$ have the same persistence exponent at $\Delta=0$, but the rate of change of the time exponent with $\Delta$ is increased as $D$ is decreased.

The linear relationship between $\theta_S$ and $\Delta$ suggests the persistence probability has a scaling relationship of 
\begin{equation}
P(\tau) \sim \tau^{- \theta_S (\Delta)} \sim \tau^{- \left(\alpha \Delta + \beta \right)} \qquad  \left(\tau \ll \tau_\times \right)
\label{eqn:small_t_scaling}
\end{equation}
in the small $\tau$ regime, where $\alpha$ is the coefficient related to the $\Delta$ dependence, and $\beta$ is the persistence exponent when $\Delta=0$.
Table~\ref{tab:theta_s_D_comparison} shows values for $\alpha$ and $\beta$ for a range of $D$.
\begin{table}[ht]
\begin{tabular}{|c|c|c|}
\hline
$\delta$-correlated & $\alpha$ & $\beta$ \\
noise strength ($D$) & & \\ 
\hline
$4.7 \times 10^6$ & 0.0971 & 0.5030 \\
$4.7 \times 10^7$ & 0.0315 & 0.5026 \\
$4.7 \times 10^8$ & 0.0107 & 0.4953 \\
$4.7 \times 10^9$ & 0.0028 & 0.5088 \\
$4.7 \times 10^{10}$ & 0.0008 & 0.5090 \\
\hline
\end{tabular}
\caption[Linear fit parameters for $\theta_S$.]{\textbf{Linear fit parameters for $\theta_S$.}
The $\alpha$ parameter is dependent on the noise strength, whereas $\beta$ is independent of the noise.}
\label{tab:theta_s_D_comparison}
\end{table}
The value of $\beta$ is constant for all $D$, but $\alpha$ is clearly dependent on $D$.
\begin{figure}[htbp!]
	\centering
	\includegraphics[width=0.45\textwidth]{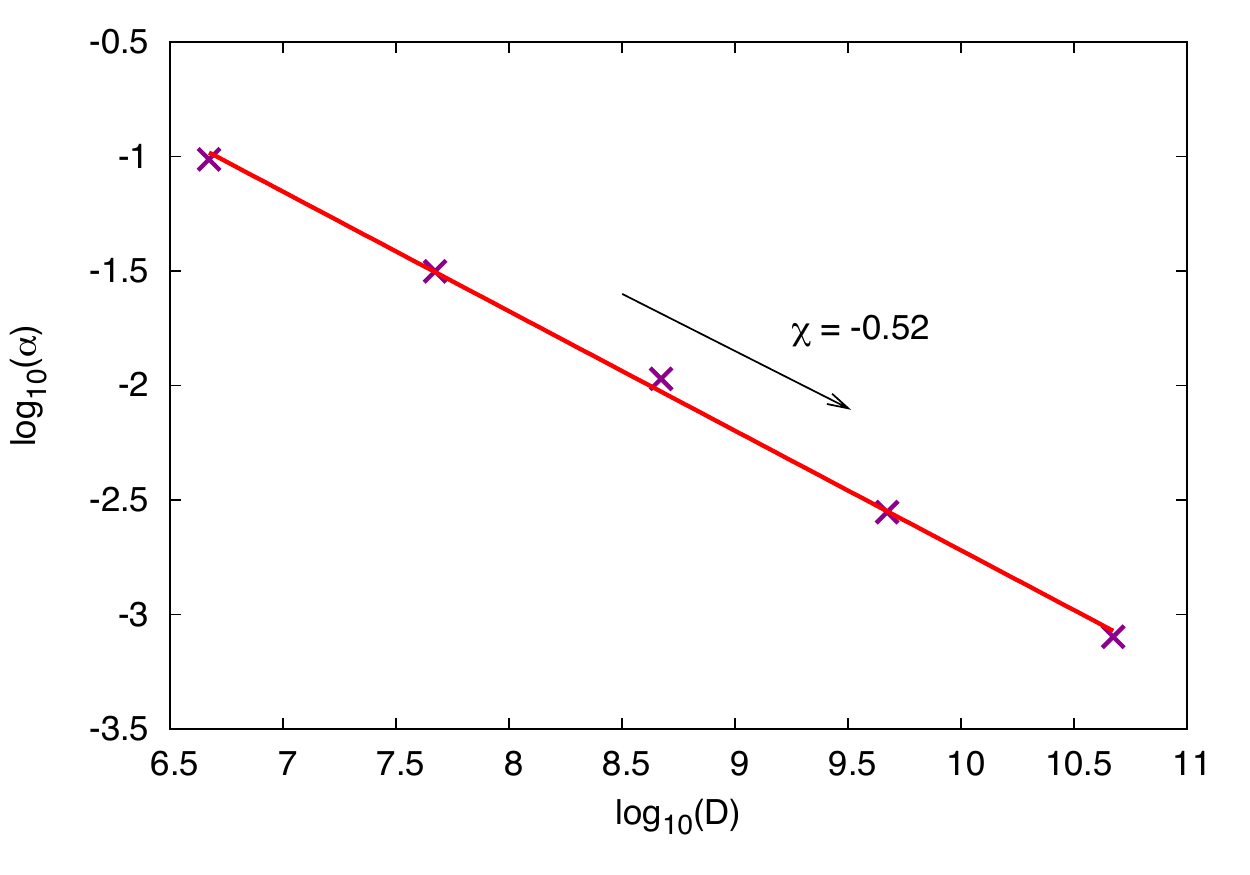}
	\caption[(Color online) The $\alpha$ dependence on $D$.]{(Color online) \textbf{The $\alpha$ dependence on $D$.} 
	Linear fit (solid line) for the time exponent parameter $\alpha$ against $D$ in the log:log scale.
	$\chi$ shows the gradient of the fitted line.
	}
	\label{fig:alpha_D}
\end{figure}
Plotting $\alpha$ against $D$ in the log:log scale (fig.~\ref{fig:alpha_D}) reveals a linear relationship that leads to a power law: $\alpha \sim D^{\chi}$.
The fitted solid line in figure~\ref{fig:alpha_D} suggests a value of $\chi \approx -0.52$; in other words, $\alpha \sim \frac{1}{\sqrt{D}}$ which when combined with the definition of the exponent $\alpha$ gives us the noise scaling of the survival probability as $P(\tau) \sim \tau^{-(\frac{\Delta}{\sqrt{D}}+\beta)}$.  Mathematically, this implies the existence of a well-defined universality class that is a function of the of the noise strength and the bond length $Delta$, that also means that numerical simulations could better use the redefined noise amplitude $\frac{D}{\Delta^2}$ instead of $D$. One must here be aware of a realistic biological constraint in that higher noise corresponds to larger thermal fluctuations and hence will be capped at some finite value.

The table above suggests that for $\Delta=0$, the survival probability is independent of the noise strength (since $\beta$ is noise independent always converging to the value 0.5) as it should be for Brownian motion. 
However, for all other values of $\Delta$, there is a competition between the free energy and noise terms that eventually determines the effective number of datapoints to be obtained numerically (the plot for $D=4.7 \times {10}^8$ in Fig. \ref{fig:average_cases_per_system} is instructive here). 
In a way, this suggests the limit of simulation accuracy in analyzing the probabilistic persistence data.
So, increasing the thermal noise leads to a greater range of statistics for increasing $\Delta$, but does not alter the time exponent relating to $\Delta=0$ nm, that corresponds to the glycocalyx length used in the linear stability analysis.
The range of $\tau$ where the scaling relation in equation~\eqref{eqn:small_t_scaling} holds is in the order of seconds and the range decreases steadily as $\Delta$ increases.

\subsection{The Large $\tau$ Phase: The \enquote{Signaling} regime}
\label{sec:results:large_t_phase}
The large $\tau$ phase represents the regime where infrequent longer lasting receptor:ligand bonds exist. This is the phase characterized by large TCR:pMHC bond half lives facilitating intra-cellular signaling required for T cell activation.
The persistence time of these bonds are in the order of tens of seconds and therefore sufficient for signaling pathways that lead to elevated levels of intracellular Ca$^{2+}$.

\begin{figure}[htbp!]
	\centering
	\includegraphics[width=0.45\textwidth]{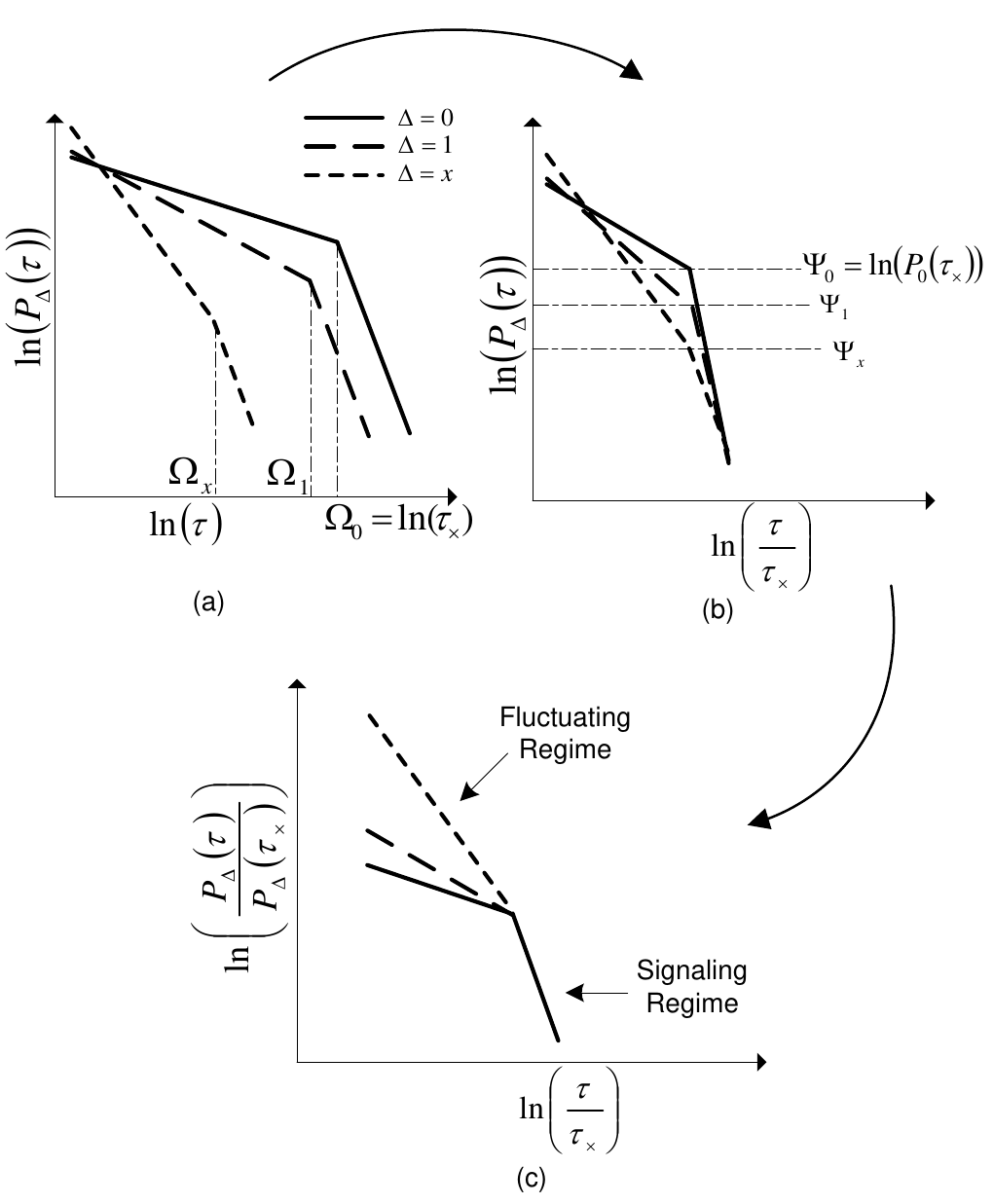}
	\caption[Schematic illustration of the time-dependent rescaling used in the large $\tau$ regime.]{\textbf{Schematic illustration of the time-dependent rescaling used in the large $\tau$ regime.}  
	Figure~\ref{fig:large_t_collapse} (a) we rescale the time dimension to ensure the crossover point between the two phases occurs at the same time for each $\Delta$, then
	figure~\ref{fig:large_t_collapse} (b) shows the rescaling step in the direction of the probability density, that collapses the large $\tau$ regime for the different $\Delta$ curves.
	Figure~\ref{fig:large_t_collapse} (c) shows the final data collapse for the rescaled curves for $\tau \ge \tau_\times$.}
	  \label{fig:large_t_collapse}
\end{figure}

\begin{figure}[htbp!]
	\centering
	\subfigure[]{
		\includegraphics[width=0.45\textwidth]{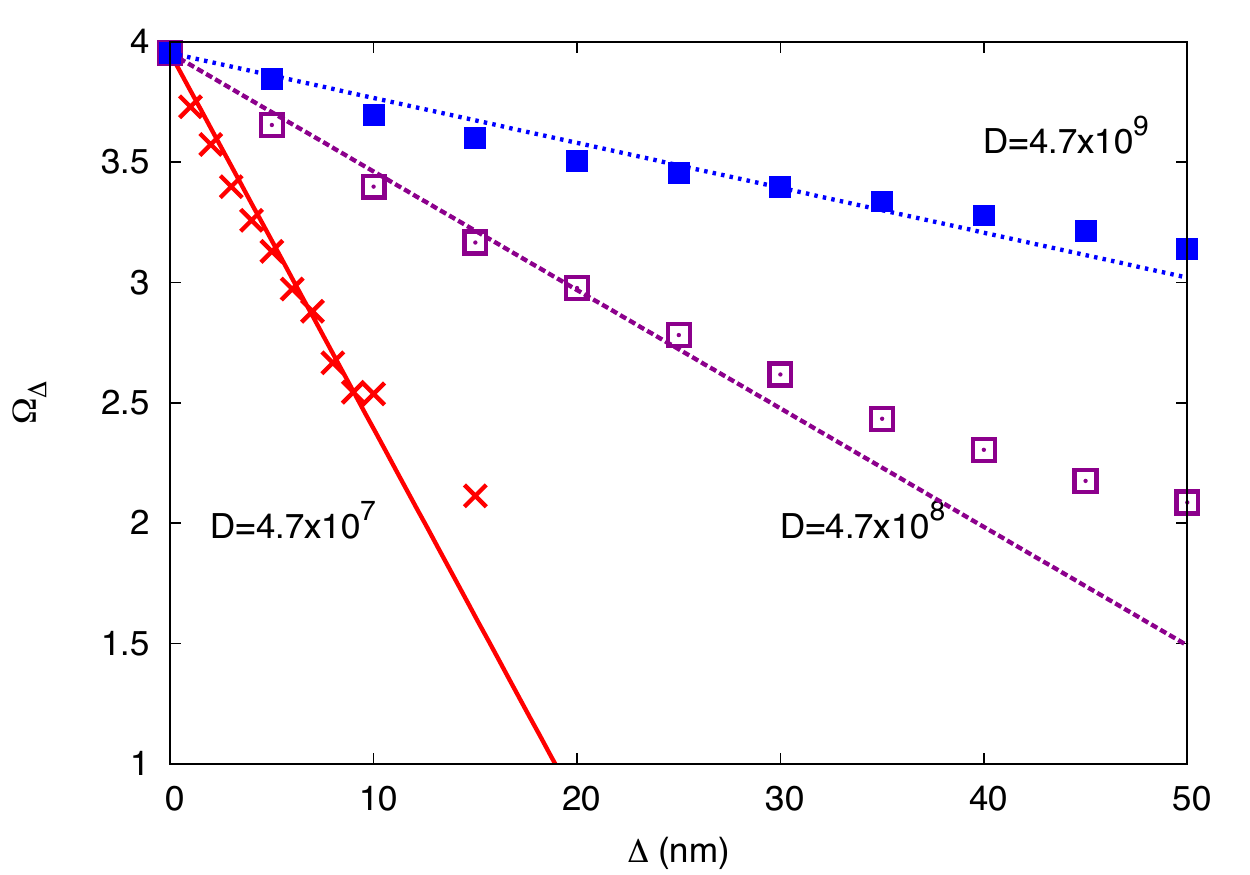}
		\label{fig:omega_vs_delta}
	}
	\\
	\subfigure[]{
		\includegraphics[width=0.45\textwidth]{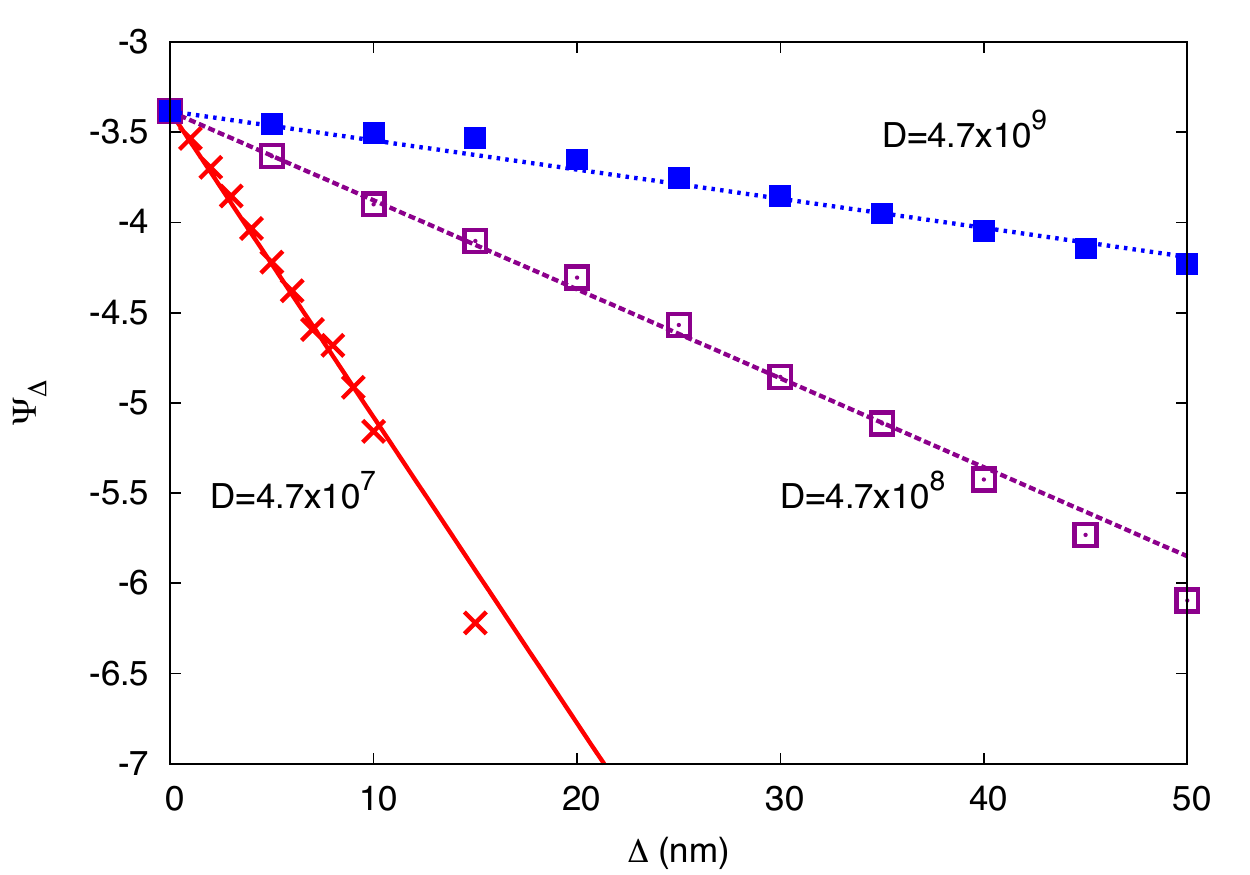}
		 \label{fig:rho_vs_delta}
	}
	\\
	\subfigure[The collapsed large $\tau$ regime.]{
		\includegraphics[width=0.49\textwidth]{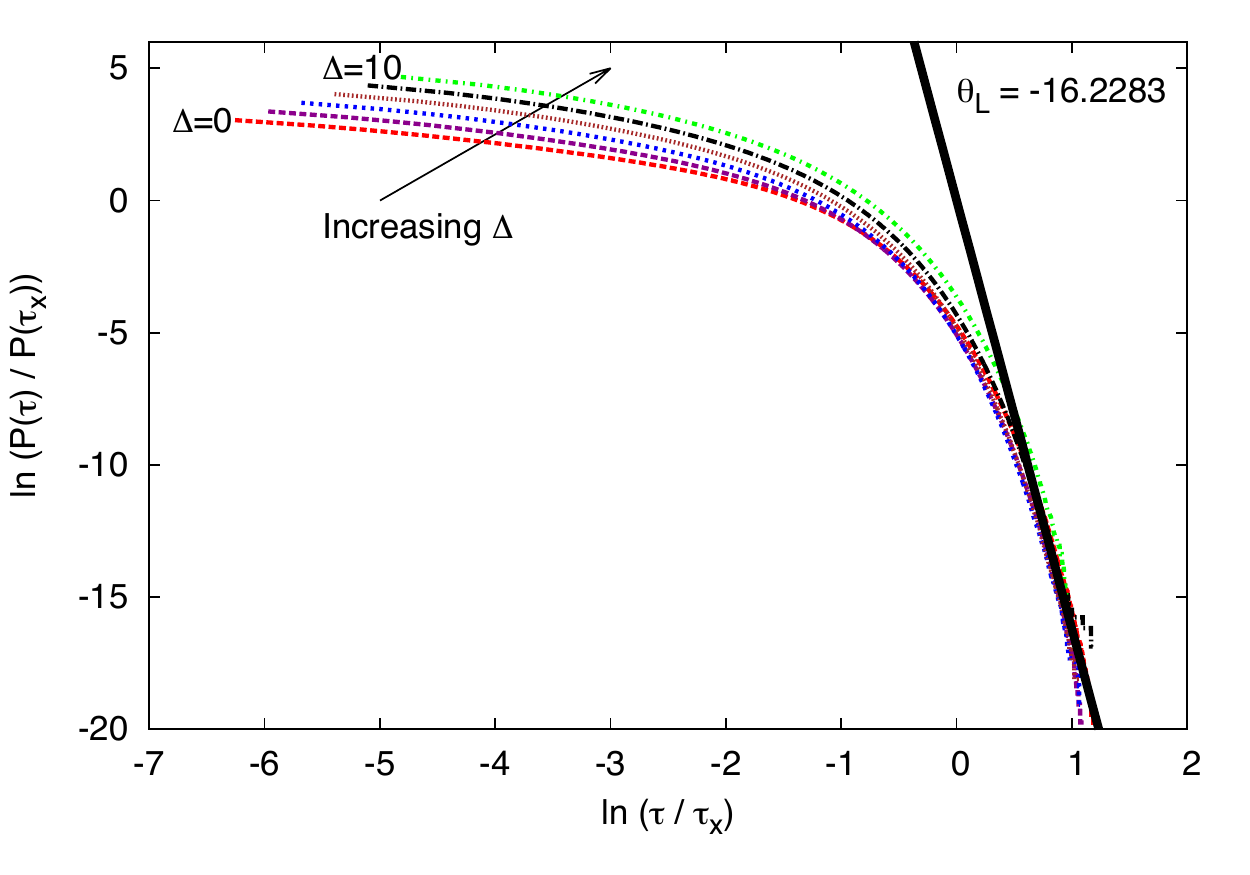}
		\label{fig:collapsed_large_t}
	}
	\caption[(Color online) The rescaling functions, $\Omega$ and $\Psi$, along with the rescaled large $\tau$ regime.]{(Color online) \textbf{The rescaling functions, $\Omega$ and $\Psi$, along with the rescaled  large $\tau$ regime.}
	Figure~\ref{fig:omega_vs_delta} shows the time dimension rescaling function $\Omega$ plotted against $\Delta$ and figure~\ref{fig:rho_vs_delta} shows the probability dimension rescaling function $\Psi$ plotted against $\Delta$.
	Figure~\ref{fig:collapsed_large_t} shows the rescaled simulation results, where the data collapse in the large $\tau$ regime can be observed.
	The solid line shows is fitted to the large $\tau$ portion and the gradient $\theta$ is shown.}
	\label{fig:rescaling_functions}
\end{figure}

Analysis of the $\theta_L$ values for different $\Delta$ reveals a consistent decay rate of $\theta_L\sim16$.
By rescaling the different curves for each $\Delta$ the large $\tau$ regime can be collapsed on to a single universal curve.
Figure~\ref{fig:large_t_collapse} shows the rescaling steps taken in the log:log scale.
We use the crossover time for a given $\Delta$, $\Omega_\Delta = \ln \left(\tau_\times \right)$, as a hard cut-off between the small and large $\tau$ regimes (fig.~\ref{fig:large_t_collapse}(a)).
$\Omega$ is plotted against $\Delta$ in figure~\ref{fig:omega_vs_delta} for three different noise strengths, where we find a near linear relationship for small $\Delta$.
As values of $\Delta$ are reached where the statistics drop off, this linear relationship begins to be questionable, that may or may not be answered with increased statistics but we do not consider it here.

We fit a straight line through the sample crossover points, $\Omega_\Delta$, for a given noise strength.
The linear relationships in figure~\ref{fig:omega_vs_delta} (solid lines) leads to the expression $\Omega=\log(\tau_\times)=-\omega \Delta + c_1$.
Similar to the analysis conducted for the small $\tau$ regime, the $\omega$ coefficient is dependent on the noise strength, but the $\Delta=0$ case is independent of noise strength.
The crossover time can be then be expressed as the exponential
\begin{IEEEeqnarray}{rCl}
\tau_\times & \approx &  e^{-\omega(D) \Delta + c_1}
\label{eqn:tau_omega}
\end{IEEEeqnarray}
Then, rescaling the time dimension using the new time variable $u=\ln(\tau / \tau_\times )$ ensures the phase transition occurs at the same point in the time dimension for all values of $\Delta$ (fig.~\ref{fig:large_t_collapse}(b)).
Similarly, rescaling in the direction of the survival probability as shown in figures~\ref{fig:large_t_collapse}(b) and~\ref{fig:rho_vs_delta}, we collapse the large $\tau$ regimes for each $\Delta$ on to a single curve.
Again, we use a linear fit giving leading to
\begin{IEEEeqnarray}{rCl}
P \left( \tau_\times \right) & \approx &  e^{-\psi(D) \Delta + c_2}
\label{eqn:p_rho}
\end{IEEEeqnarray}
Equations \eqref{eqn:tau_omega} and \eqref{eqn:p_rho} remain impervious to changes in the noise strength for $\Delta=0$, thereby indicating the existence of an universal survival probability for $\Delta=0$.
This result has a remarkably important biological connotation.
While all other persistence statistics are shown to be scale dependent, the cross-over time regime is a scale independent dynamics, suggesting that parameter values could always be optimized to attain non-equilibrium dynamics. 

Using the new time variable  $u$ we can map the survival probability in the large $\tau$ regime on to the universal problem
\begin{equation}
f(u) = - \theta' u \qquad 
\label{eqn:large_t_universal}
\end{equation}
that is valid for $u \ge 0$ corresponding to $\tau \ge \tau_\times$.
Using $f(u)=\ln \left( P (\tau) / P(\tau_\times) \right)$ in equation~\eqref{eqn:large_t_universal} we can rearrange to get the survival probability scaling
\begin{equation}
P(\tau) \sim \tau^{-\theta'} 
\end{equation}
where $\theta'=\theta_L \sim 16$ and from equations~\eqref{eqn:tau_omega} and~\eqref{eqn:p_rho} the magnitude of the survival probability for a given $\Delta$ and noise strength is given by
\begin{IEEEeqnarray}{rCl}
\frac{P(\tau_\times)}{\tau_\times^{-\theta'}}  & = &  e^{-\left( \psi(D) + \omega(D) \theta' \right) \Delta + c_2 + \theta' c_1}
\end{IEEEeqnarray}
confirming the universal persistence probability for $\Delta=0$, for all $D$.
Finally, figure~\ref{fig:collapsed_large_t} shows the collapsed large $\tau$ regime for the simulation results using a noise strength of $D=4.7 \times 10^8$, that is representative of the amplitude required to stimulate dynamics on a scale required for the TCR:pMHC and ICAM-1:LFA-1 bonds.
The data collapse suggests an universal scaling regime for the decay rate of the persistence probability during longer bond attachments.

\subsection{Extremal Value Statistics}
\label{sec:results:extremal_value_statistics}

The average time persistence calculated using equation~\eqref{eqn:t_average} is a monotonically decreasing function as the threshold increases~\cite{bib:bush_2014}.
The fluctuations due to thermal noise lead to rapid crossings of the threshold as the separation distance moves from a close contact phase to one of separation and vice-versa.
Here we look at the extent to which these rapid crossings contribute to the average time persistence for the bonds.

We term the time persistence realizations due to rapid fluctuations as extremal values, where the bond life is extremely short.
Table~\ref{tab:max_likelihood} shows the normalized frequency distribution of $t_{\Delta}$ for a range of $\Delta$.
\begin{table}[ht]
\centering
\begin{tabular}{|c|c|c|c|c|}
\hline
 & \multicolumn{3}{|c|}{First Passage Prob Time} & \\
$\Delta$ (nm) & $\Delta t$ & $2 \Delta t$ & $3 \Delta t$ & Total \\
\hline
0 & 0.2945 & 0.1315 & 0.0789 & 0.5048 \\
5 & 0.3072 & 0.1366 & 0.0818 & 0.5257 \\
10 & 0.3201 & 0.1417 & 0.0846 & 0.5465 \\
15 & 0.3333 & 0.1467 & 0.0873 & 0.5672 \\
20 & 0.3466 & 0.1515 & 0.0897 & 0.5878 \\
25 & 0.3601 & 0.1562 & 0.0919 & 0.6082 \\
30 & 0.3738 & 0.1607 & 0.0939 & 0.6283 \\
\hline
\end{tabular}
\caption{The relative frequency for occurrences of transient bonding.
The number of bonds that have a duration of 3 discrete time step lengths or less account for 50\% or more of the statistics.}
\label{tab:max_likelihood}
\end{table}

\begin{figure}[htbp!]
	\centering
	\includegraphics[width=0.48\textwidth]{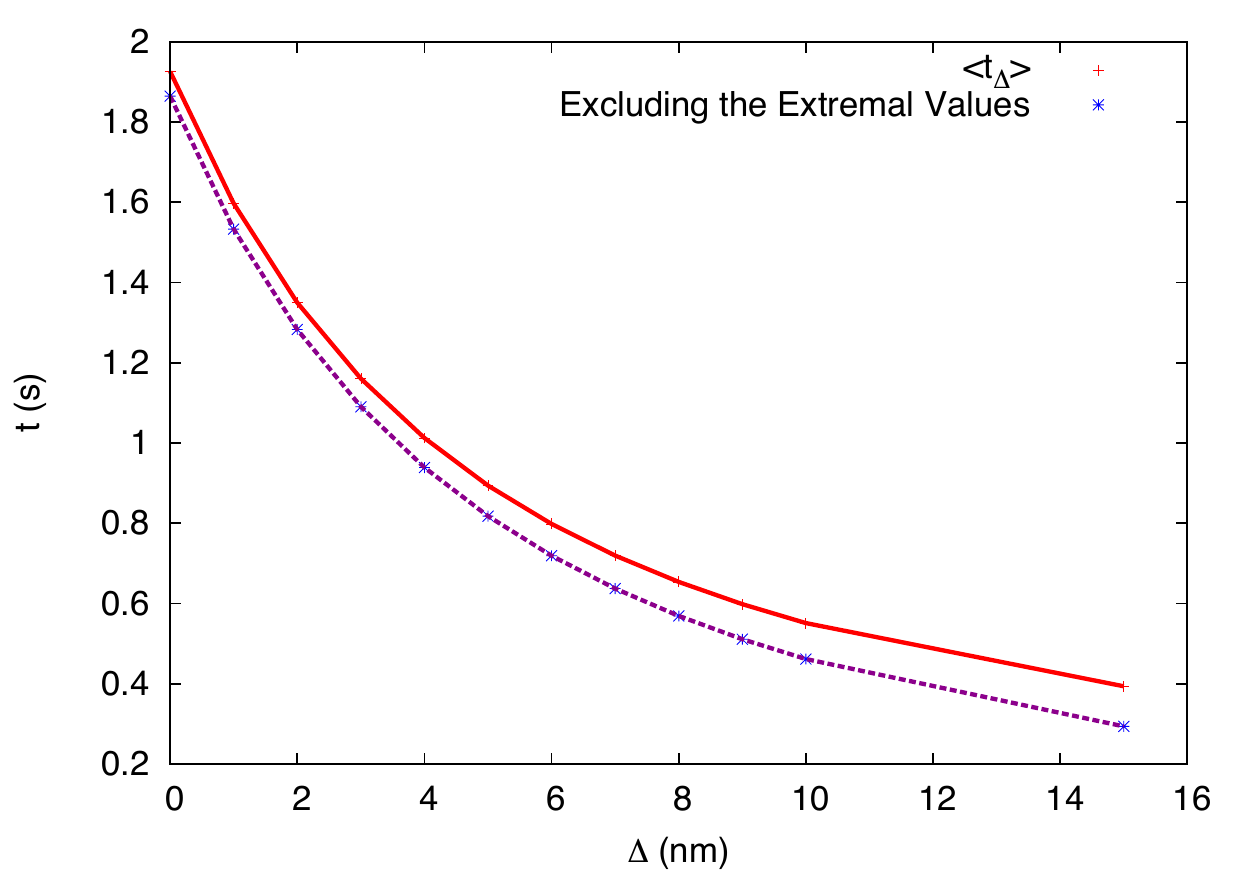}
	\caption[(Color online) The $<t_{\Delta}>$ time persistence against $\Delta$]{(Color online) \textbf{The $\left< t_{\Delta} \right>$ time persistence against $\Delta$}. The average time persistence plotted against $\Delta$ (solid line), along with the average time persistence excluding the extremal events (dotted line).}
	\label{fig:t_plus_avg_vs_delta}
\end{figure}
At least 50\% of the statistics for each $\Delta$ are accounted for in the first three discrete time steps.
That is, at least 50\% of bonds forming will disassociate in a fraction of a second.

From our computer simulations we calculate approximations for the $\Delta$ dependent probability density function $W_\Delta$, equation~\eqref{eqn:probability_density}, and this is used to calculate the average time persistence.
To assess the contribution of the extremal values to the total average we introduce a lower cut-off $v \ge 0$ to the first moment calculation
\begin{equation}
\left< t_{\Delta} \right> = \int_v^{\infty} t~W_\Delta \left( t \right) dt
\label{eqn:t_average_cutoff}
\end{equation}
which is used to exclude the extremal values from the calculation.
In our case, we set $v=3 \Delta t$ to exclude the statistics set out in table~\ref{tab:max_likelihood}.
Figure~\ref{fig:t_plus_avg_vs_delta} shows $\left< t_\Delta \right>$ against $\Delta$ (solid line) along with the corresponding average time persistence excluding the rapid fluctuations (dotted line), created using a $D=4.7 \times 10^8$ thermal noise strength.

For small $\Delta$ the extremal values have very little contribution to the total average calculation, however the contribution steadily increases as $\Delta$ increases. 

\section{Conclusions}
The analysis presented here has the following immunological implications.
First, figure~\ref{fig:t_plus_avg_vs_delta} clearly shows that extremal value statistics do not contribute greatly to the average time persistence in the linear stability region. 
Therefore, rapidly fluctuating membrane dynamics following high energy dissipation has little effect on the statistics and the survival probability profile may be well described excluding large amplitude fluctuations.
We aim to use this result in future work to model the mature immunological synapse (in the nonlinear regime).

Next, the survival probability distribution plotted in figure~\ref{fig:log_pdf_function_method} shows two distinct phases, corresponding to two separate time exponents.
The small-$\tau$ regime corresponds to a rapidly fluctuating membrane about the threshold separation distance, where transient bonds are associating and disassociating, while the large $\tau$ regime corresponds to the bonds that persist for time periods of the order of magnitude required for downstream signaling that leads to cell activation.
The small-$\tau$ regime has a time exponent that is dependent on both the threshold bond length and the noise amplitude. As shown in figure~\ref{fig:alpha_D}, the exponent $alpha$ scales with the noise amplitude $D$ thereby defining a \enquote{diffusive universality class} between $\alpha$ and $D$ that quantifies into the rescaling of the noise amplitude as $\frac{D}{\Delta^2}$. From the perspective of a biologist, the above result implies that if the system is calibrated with respect to the dimensionless variable $\frac{D}{\Delta^2}$ instead of the two variables $D$ and $\Delta$ independently, the signaling domain can be directly identified from the distribution of \enquote{time patch} sizes as is shown through the estimation of a single valued $\theta$ in figure~\ref{fig:collapsed_large_t}. For a signaling setup involving multiple coreceptor molecules with varying bond lengths, the above analysis could enable the prediction of the start of the signaling regime for a fixed noise input based on the results presented here.
Fixing the thermal noise leads to a linear dependency between the time exponent and the bond length.
However, the small-$\tau$ regime has a universal time exponent that is defined by $\frac{D}{\Delta^2}$, indicating a constant dissociation rate ($\text{k}_{\text{off}}$ as in \cite{bib:burroughs_2002,bib:mckeithan_1995}), that appears in the parameter $\lambda$ in the base model, which can be used for the bond lengths considered (15 - 45nm) in this regime.

\end{document}